\documentclass[pre,amsmath,amssymb,twocolumn,showpacs,showkeys]{revtex4}
\usepackage{graphicx}
\usepackage{dcolumn}
\usepackage{bm}
\usepackage{subfigure}
\usepackage{epsfig}
\usepackage{color}

\newcommand{\comment}[1]{}
\begin{document}

\preprint{APS/123-QED}

\title{Quantum Transport through Hierarchical Structures}
\author{S. Boettcher}
\email{sboettc@emory.edu}
\homepage{http://www.physics.emory.edu/faculty/boettcher/}
\affiliation{Dept. of Physics, Emory University, Atlanta, GA
30322, USA}
\author{C. Varghese}
\affiliation{Dept. of Physics \& Astronomy, Mississippi State University,
 Mississippi State, MS 39762, USA}
\author{M. A. Novotny}
\affiliation{Dept. of Physics and $HPC^2$ Center for
Computational
    Sciences, P.O. Box 5167, Mississippi State University,
    Mississippi State, MS, 39762, USA}
\date{\today}

\begin{abstract}
The transport of quantum electrons through hierarchical lattices is of interest because 
such lattices have some properties of both regular lattices and random systems.  
We calculate the electron transmission as a function of energy in the tight binding 
approximation for two related Hanoi networks. HN3 is a Hanoi network with every site 
having three bonds.  HN5 has additional bonds added to HN3 to make the average number of 
bonds per site equal to five.  We present a renormalization group approach 
to solve the matrix equation 
involved in this quantum transport calculation. We observe band gaps in HN3, while no 
such band gaps are observed in linear networks or in HN5.
\end{abstract}

\maketitle

\section{Introduction}
Understanding and controlling the transport of electrons is central to the
operation of all electrical and electronic devices \cite{switch}.  In many cases of
interest in nanomaterials, the electron transport is coherent, and therefore
must be analyzed using the Schr{\"o}dinger equation.  Interference effects can
then lead to a metal-insulator transition and Anderson localization
\cite{anderson}.  Even more than fifty years after Anderson's publication
of his celebrated paper, the effect remains an active area of study
\cite{Lage2009,monthus,montusRG,caliskan,islam,Liu2007,Ostr2006}.  The main
goal is to calculate the transmission probability as a function of energy, $E$, for
an incoming electron, i.e. the probability that an electron that comes in
from $x=-\infty$ can be observed at $x=+\infty$.

The starting point to quantum calculations of (spinless) electrons through
a material is often a tight-binding model \cite{datta}.  In such a
model each node can be considered an atom, the on-site energy at a node is associated with
a potential energy at the site, and there is a hopping term which comes from
discretization of the kinetic energy term of the Schr{\"o}dinger equation \cite{datta}.
The electron transport calculation via the Schr{\"o}dinger equation thus reduces to the
solution of an infinite matrix equation.  The solution of the matrix equation is often
accomplished using a Green's function method 
\cite{andrade,caliskan,datta,datta3,young,pouthier,ryndyk}.
In this paper we instead use an ansatz approach introduced by Daboul, Chang, and Aharony
\cite{daboul}, which is simpler to describe at an undergraduate level \cite{Solomon2010}.
The ansatz reduces the size of the matrix equation to that of the number of tight-binding
cites in the scattering volume, plus one for both the incoming lead and the outgoing lead.
This approach has been used by other authors \cite{islam,cuansing}.
We find that the ansatz approach is particularly well suited to our calculations of
transport through hierarchical structures.
To perform the calculation we construct a decimation Renormalization Group (RG)
procedure to reduce the number of sites of the hierarchical structure.  Our RG is
related to that utilized by others for tight-binding models 
\cite{aoki,Banav1983,Niu1986,Mama2003}.
However, our RG has been explicitly constructed for the calculation of the transmission 
probabilities.  We find that our RG procedure greatly simplifies the calculation, albeit only 
for certain select networks that have hierarchical structures. Although our RG
does not significantly simplify the calculation of associated wavefunctions, we 
nevertheless give a recipe for the RG calculation of the wavefunctions.

The specific models we solve here for quantum transport are motivated by four
considerations.  One is to understand how hierarchical models, in particular the
Hanoi networks \cite{SWPRL}, affect quantum transport.  The second is that such
hierarchical models provide an intermediate between regular lattices and ones
that have a small-world property \cite{Watts}, and would be of interest to
understand quantum transport of nanomaterials that have a small-world property
\cite{caliskan,Novotny2004,Novotny2005,Zhang2006,yancey2009}.  The third is that
often phase transitions such as the metal-insulator transition or a ferromagnetic
transition have some universal quantities that depend only on the dimension.
Hierarchical models then sometimes provide insight into how these
universal quantities behave as a function of the dimension
\cite{Gefen1980,Banav1983,LeGuillou1987,Novotny1992,Novotny1993}.
Lastly, experimental realizations of hierarchical materials may possess
novel physical properties \cite{Lopes2001,Boker2004,Lin2005,Ravind2005,Chen2008}.

The hierarchical models we study are the Hanoi networks HN3 and HN5.  These networks have 
particularly interesting properties.  First, they are both planar, and consequently 
could be experimentally constructed on a surface.  Second, both networks have typical 
\lq paths' (defined precisely in Sec.~\ref{sec:Graph-Structure}) that grow more slowly 
with system size than do paths in regular lattices.  Finally, Anderson localization is 
associated with randomness in the system, while randomness in the Hanoi networks depends 
on the scale.  For example, for HN3 locally every site has three bonds connecting it with 
other sites, choosing sites from HN3 at random the connections to other sites seem random, 
but these connections are actually from a hierarchical arrangement and hence at the 
larger scale there is regularity to the network.  Therefore studying electron transport and 
Anderson localization in these lattices is of interest.  

In Section~\ref{sec:Graph-Structure} we provide a brief description of the Hanoi networks.
For transport properties, these networks can be connected to the leads in many ways, but
we choose to present results only for symmetric linear and symmetric ring lead attachments.
In Section~\ref{MatrixRG} we develop the RG equations for calculating the transmission
through these networks, with details of the RG presented in Appendix~\ref{Sec:AppA}.
In Section~\ref{analysis} we analyze these RG equations for these networks.
This involves iterating the RG until the system is comprised only of a few lattice points,
and these small lattice solutions are presented in Appendix~\ref{Sec:AppB}.
Section~\ref{conclusion} contains our conclusions and a discussion of our results.
The appendices contain the basic
matrix algebra used to develop the RG equations.

\section{Network Structure\label{sec:Graph-Structure}}
\begin{figure}
\includegraphics[bb=100bp 0bp 800bp 680bp,clip,scale=0.3]{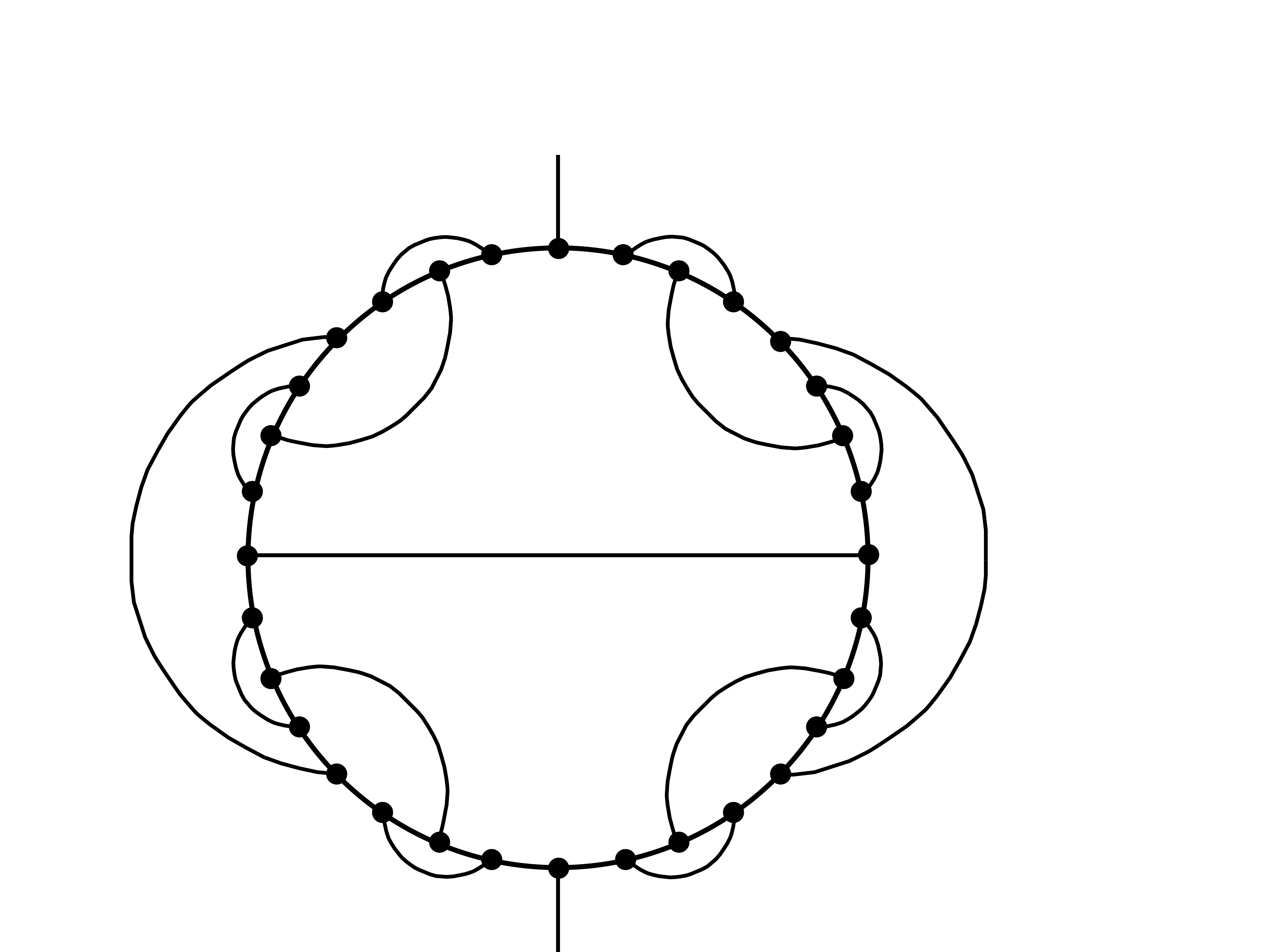}
\caption{\label{fig:3hanoi}
Depiction of the 3-regular network HN3 with a one-dimensional
periodic backbone forming a ring, here with $k=5$. The top and bottom sites 
$n=0$ and $n=2^{k-1}$ require special treatment and are connected
to external leads. With these connections, the entire network
becomes 3-regular. Note that the graph is planar.}
\end{figure}

\begin{figure}
\includegraphics[bb=90bp 100bp 400bp 700bp,clip,angle=-90,scale=0.4]{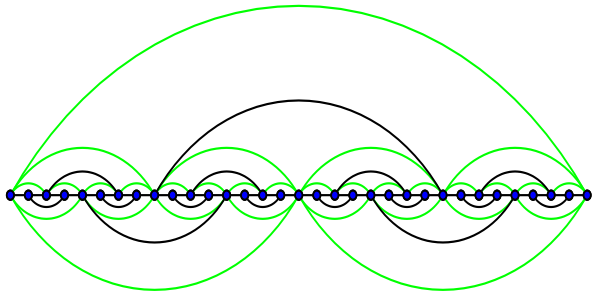}
\caption{\label{fig:5hanoi} (Color online)  Depiction of the planar network HN5. 
Black lines demark the original HN3 
structure, the green-shaded lines are added to make HN5. 
Note that sites on the lowest level of the hierarchy have degree 3, then degree
5, 7, $\cdots$, making up a fraction of 1/2, 1/4, 1/8, $\cdots$, of all sites,
thereby making for an average degree 5 of this network. }
\end{figure}

\begin{figure}
\includegraphics[bb=0bp 100bp 1024bp 500bp,clip,scale=0.23]{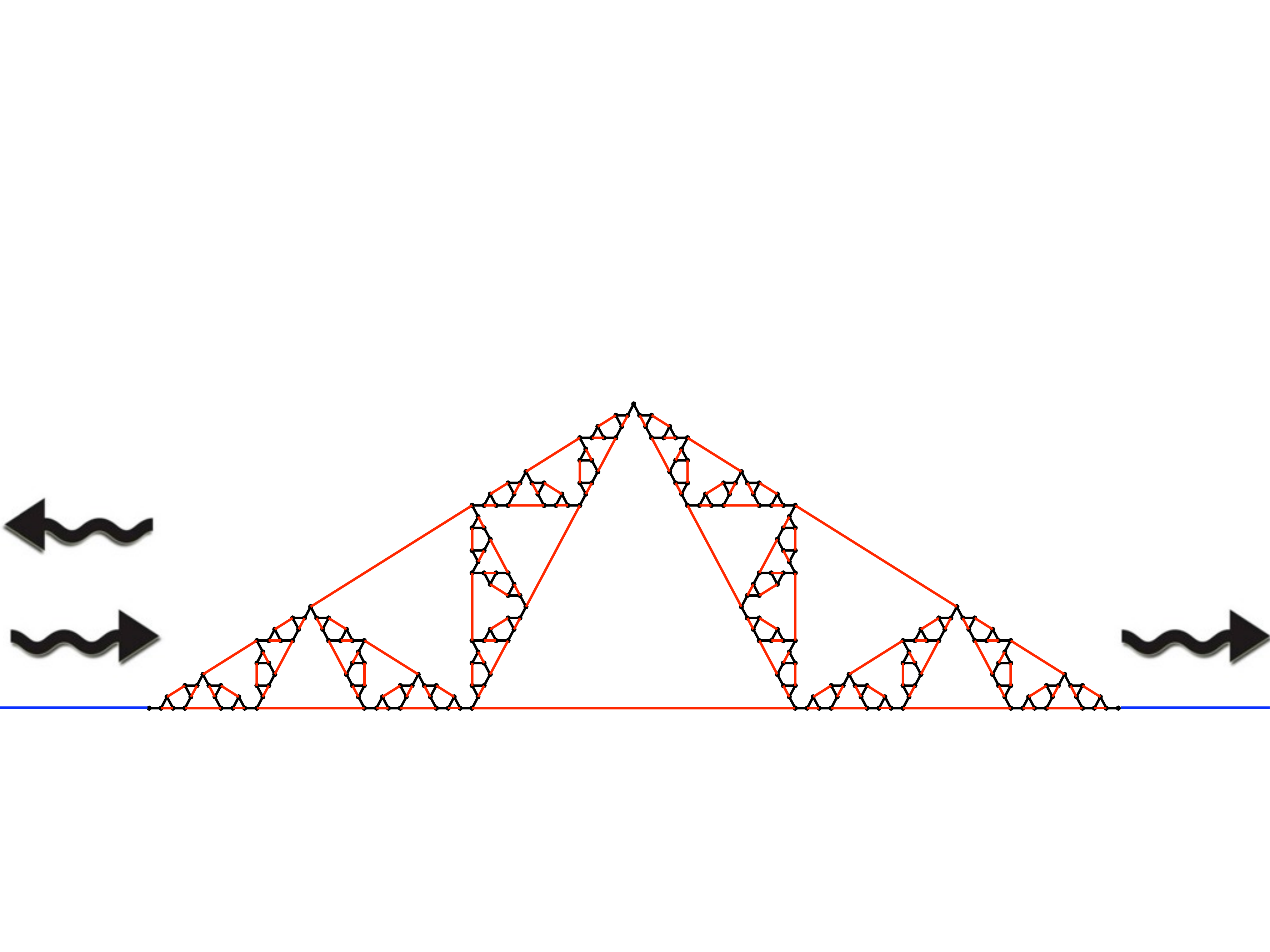}
\caption{\label{fig:HN3scatterKoch} (Color online) Scattering a quantum electron 
off a linear version of HN3, here drawn as a branched Koch curve. The incoming
electron on the left gets scattered into an outgoing transmitted portion
(right) and reflected portion (left) on the attached external leads (blue-shaded).
In this form of HN3, the one-dimensional backbone is marked by black
links while the small-world links are shaded in red. Note, for instance,
that the shortest end-to-end path here is the baseline of the Koch
curve.}
\end{figure}

Each of the networks considered in this paper possess a
simple geometric backbone, a one-dimensional line of $N=2^{k}$
sites formed into a ring as depicted in Fig.~\ref{fig:3hanoi} for 
HN3 and $k=5$. Alternatively, we can connect the network 
to the incoming and outgoing leads in a linear arrangement 
with $2^k+1$ sites, as depicted in Fig.~\ref{fig:5hanoi} and 
\ref{fig:HN3scatterKoch}.  
Each site is at least connected to its nearest
neighbor left and right on the backbone. For consistency, we
call the ordinary one-dimensional ring HN2 (for Hanoi Network
of degree 2).  For example, HN2 is the linear lattice in 
Fig.~\ref{fig:HN3scatterKoch} formed by only the black bonds. 

To generate the small-world hierarchy in these networks, 
consider parameterizing any number $n<N$ (except for
0) \emph{uniquely} in terms of two other integers $(i,j)$,
$i\geq1$ and $1\leq j\leq2^{k-i}$, via
\begin{eqnarray}
n & = & 2^{i-1}\left(2j-1\right).
\label{eq:numbering}
\end{eqnarray}
Here, $i$ denotes the level in the hierarchy whereas $j$ labels
consecutive sites within each hierarchy. To generate the
network HN3, we connect each site $n=2^{i-1}(4j-3)$ also with a
long-distance neighbor $n'=2^{i-1}(4j-1)$ for $1\leq
j\leq 2^{k-i-1}$. (In the ring, if an index $n$ equals 
or exceeds the system size $N$, we assume that the site $n$ mod
$N$ is implied.) 

For the linear arrangement, the sites zero and $2^k+1$ are connected to 
the input/output leads, and site $2^{k-1}$ will not be connected to any other 
site or to the input/output leads [see, for instance, HN5 in Fig.~\ref{fig:5hanoi}].  
For the ring arrangement, the sites with number zero and $2^{k-1}$ are connected to 
the input/output leads [see Fig.~\ref{fig:3hanoi}], and 
site zero is connected to site $2^k-1$ to form the ring.  

Previously \citep{SWPRL}, it was found that the average
\lq\lq chemical path'' between sites on HN3 scales as
\begin{equation}
d^{HN3}\sim\sqrt{l}
\label{eq:3dia}
\end{equation}
with the distance $l$ along the backbone. In some ways, this
property is reminiscent of a square-lattice consisting of $N$
lattice sites with diagonal $\sim\sqrt{N}$.

While the preceding networks are of a fixed, finite degree, we
can extend HN3 in the following manner to obtain a new planar
network of average degree 5, hence called HN5, at the price of
a distribution in the degrees that is exponentially falling. In
addition to the bonds in HN3, in HN5 we also connect all even
sites to both of its nearest neighboring sites \emph{within}
the same level of the hierarchy $i>1$ in
Eq.~(\ref{eq:numbering}). The resulting network remains planar
but now sites have a hierarchy-dependent degree, as shown in
Fig.~\ref{fig:5hanoi}. To obtain the average degree, we observe
that 1/2 of all sites have degree 3, 1/4 have degree 5, 1/8
have degree 7, and so on, leading to an exponentially falling
degree distribution of ${\cal P}\left\{ \alpha=2i+1\right\}
\propto2^{-i}$ for $i=1,\>2,\>3,\cdots$. Then, the total number of bonds 
$L$ in the (linear) system of size $N=2^{k}+1$ is
\begin{eqnarray}
 2L & =2(2k-1)+ & \sum_{i=1}^{k-1}\left(2i+1\right)2^{k-i}=5\times 2^{k}-4,
\label{eq:TotalLinksHN5}
\end{eqnarray}
thus, the average degree is
\begin{eqnarray}
\left\langle \alpha\right\rangle  & = & \frac{2L}{N}\sim5.
\label{eq:averageDegreeHN5}
\end{eqnarray}

In HN5, the end-to-end distance is trivially 1, see
Fig.~\ref{fig:5hanoi}. Therefore, we define as the diameter the
largest of the shortest paths possible between any two sites,
which are typically odd-index sites furthest away from
long-distance bonds. For the $N=32$ site network depicted in
Fig.~\ref{fig:5hanoi}, for instance, that diameter is 5 as 
measured between site 3 and 19 (0 is the left-most site),
although there are many other such pairs. It is easy to show
recursively that this diameter grows strictly as
\begin{eqnarray}
d^{HN5} & = & 2\left\lfloor k/2\right\rfloor +1\sim\log_{2}N 
\label{eq:5dia}
\end{eqnarray}
with $\lfloor x\rfloor$ the integer portion of $x$.  
We have checked numerically that the \emph{average} shortest
path between any two sites also increases logarithmically
with system size $N$.

\section{Matrix RG for Hanoi Networks\label{MatrixRG}}

\begin{figure}
  \includegraphics[width=.5\textwidth, bb=0bp 0bp 494bp 537bp,clip]{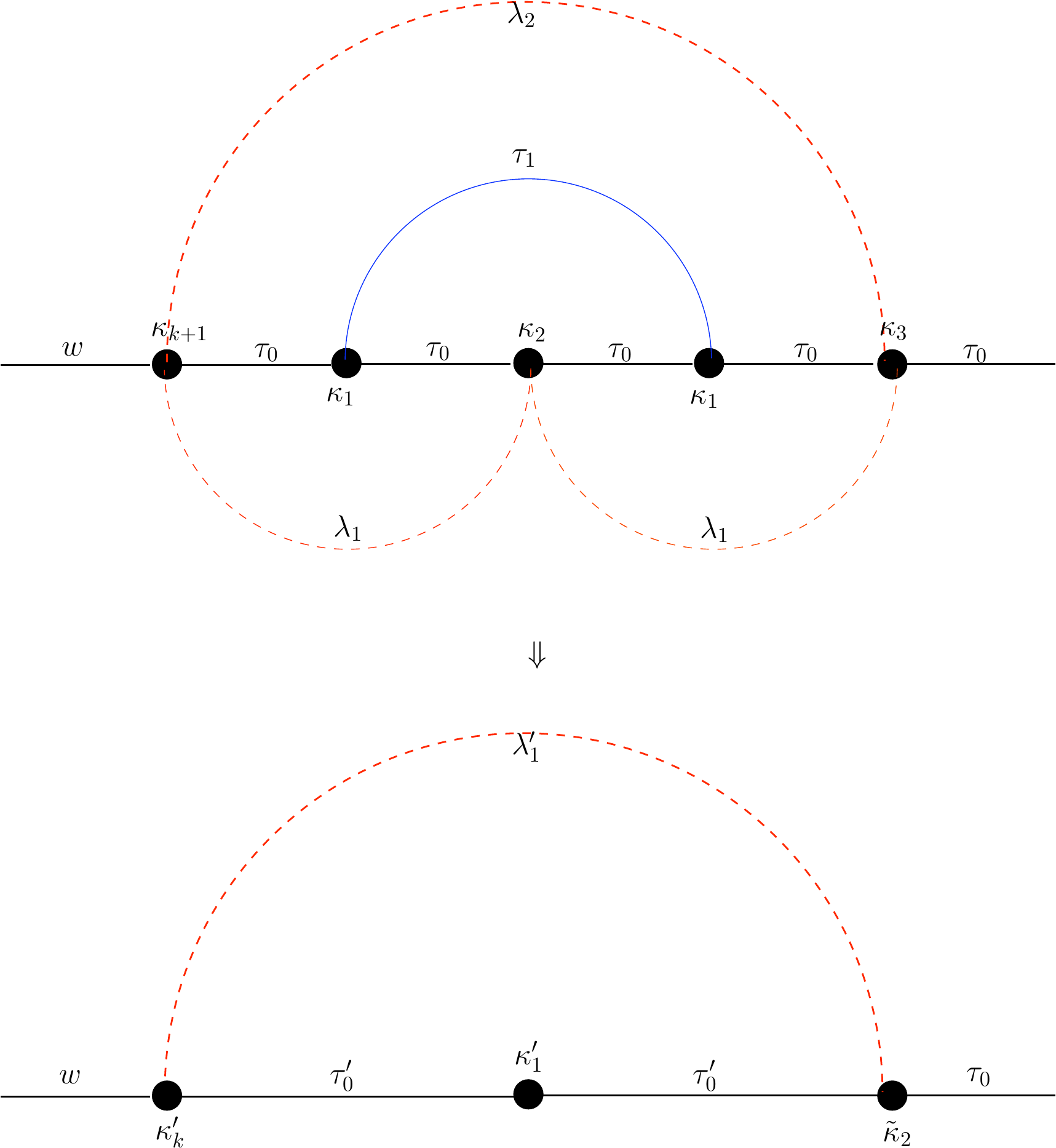}\\
  \caption{(Color online) Decimation of a block in the RG.  The two sites 
           with on-site energy $\kappa_1$ with a connecting bond of strength $\tau_1$ 
           are decimated.}
  \label{BuildingBlock}
\end{figure}
At each RG step, we decimate all the odd sites. 
Take site 0 to be at level $i =k+1$ in a linear geometry and at level $i  =k$ in a ring geometry.
As the odd sites each have only one small-world-type bond,
we can divide the network into blocks containing 5 sites and
decimate the pairs of two odd sites block by block. Let us start with the
first block which contain sites 0,1,2,3 and 4. This decimation
process for a linear geometry is shown in Fig.~\ref{BuildingBlock}. Thus,
here (see Appendix~\ref{Sec:AppA}) we have
\begin{equation}
 {\bf A} =
             \begin{pmatrix}
               \kappa_{k+1} & \lambda_{1} & \lambda_{2} \\
               \lambda_{1} & \kappa_{2} & \lambda_{1} \\
               \lambda_{2} & \lambda_{1} & \kappa_{3} \\
             \end{pmatrix} 
\end{equation}
and
\begin{eqnarray}
    {\bf D} =
                  \begin{pmatrix}
                    \kappa_{1} & \tau_{1} \\
                    \tau_{1} & \kappa_{1} \\
                  \end{pmatrix}
                ; \qquad {\bf B} =
                                            \begin{pmatrix}
                                              \tau_{0} & 0 \\
                                              \tau_{0} & \tau_{0} \\
                                              0 & \tau_{0} \\
                                            \end{pmatrix} = \tau_{0}\begin{pmatrix}
                                              1 & 0 \\
                                              1 & 1 \\
                                              0 & 1 \\
                                            \end{pmatrix} \label{AdB}
.\end{eqnarray}

After decimation,
\begin{eqnarray}
{\bf A'} =
             \begin{pmatrix}
               \kappa'_{k} & \tau'_{0} & \lambda'_{1} \\
               \tau'_{0} & \kappa'_{1} & \tau'_{0} \\
               \lambda'_{1} & \tau'_{0} & \tilde{\kappa}_{2} \\
             \end{pmatrix}
            = {\bf A} -{\bf B}{\bf D}^{-1}{\bf B}^{\rm T}
.\end{eqnarray}
After simplification, we find that
\begin{eqnarray}
\kappa'_{k} &=& \kappa_{k+1} - 
    \frac{\tau_{0}^{2}\kappa_{1}}{\kappa_{1}^{2}-\tau_{1}^{2}} \label{endsite}\\
\kappa'_{1} &=& \kappa_{2} - \frac{2\tau_{0}^{2}}{\kappa_{1} + \tau_{1}} \\
\tilde{\kappa}_{2} &=& \kappa_{3} - \frac{\tau_{0}^{2}\kappa_{1}}{\kappa_{1}^{2}-\tau_{1}^{2}} \\
\tau'_{0} &=& \lambda_{1} - \frac{\tau_{0}^{2}}{\kappa_{1} + \tau_{1}} \\
\lambda'_{1} &=& \lambda_{2} + \frac{\tau_{0}^{2}\tau_{1}}{\kappa_{1}^{2}-\tau_{1}^{2}} 
     \label{nexttonear}
.\end{eqnarray}
Here, the primed and unprimed quantities represent the $1$-st
and the $0$-th level respectively in the RG recursion. Notice
that the decimation of sites connected to site~4 [the right-most site in 
Fig.~\ref{BuildingBlock}(a)] is not complete yet and therefore its on-site 
energy will be modified further when we decimate the odd
sites of the next block which contain sites 4,5,6,7 and 8.

After the decimation of the next block we get
\begin{eqnarray}
  \kappa'_{2} &=& \tilde{\kappa}_{2} - \frac{\tau_{0}^{2}\kappa_{1}}{\kappa_{1}^{2}-\tau_{1}^{2}} = 
  \kappa_{3} - \frac{2\tau_{0}^{2}\kappa_{1}}{\kappa_{1}^{2}-\tau_{1}^{2}}
.\end{eqnarray}
Continuing the decimation block by block in this way, we find
that
\begin{eqnarray}
  \kappa'_{i} &=& \kappa_{i+1} - \frac{2\tau_{0}^{2}\kappa_{1}}{\kappa_{1}^{2}-\tau_{1}^{2}} 
    \quad \textrm{for} \quad \forall i \in \{2,\ldots,k\} \label{evensite}\\
  \tau'_{i} &=& \tau_{i+1} \quad \forall i\geq1 \label{taumatrix}\\
  \lambda'_{i} &=& \lambda_{i+1} \quad \forall i\geq2 \label{lambdamatrix}
.\end{eqnarray}
At first it appears that there are a lot of RG variables to
worry about. However most of these RG variables are
interdependent. It can be deduced from
Eqs.~(\ref{endsite},\ref{evensite}, \ref{taumatrix},\ref{lambdamatrix}) that
\begin{eqnarray}
  \tau_{1}^{(m)} &=& \tau_{m+1} \\
  \lambda_{2}^{(m)} &=& \lambda_{m+2}
  ,\end{eqnarray}
 and that the on-site energy parameter of the even sites is related to those of the odd sites as
 \begin{widetext}
  \begin{eqnarray}
  \kappa_{i}^{(m)} &=& 
  \kappa_{1}^{(m)} + \kappa_{m+i} - \kappa_{m+1} - 2(\lambda_{1}^{(m)} - \lambda_{m+1}) 
  \quad \textrm{for} \quad \forall i \in \{2,\ldots,k-m\} \label{aries}
  \end{eqnarray}
and that specifically for a linear geometry, the on-site energy parameter of the end sites
  \begin{eqnarray}
  \kappa_{k+1-m}^{(m)} &=& \kappa_{k+1} + (\kappa_{2}^{(m)} - \kappa_{m+2})/2 \quad \textrm{for} 
  \quad \forall m \in \{0,\ldots,k-2\} \label{cancer}
.\end{eqnarray}
Thus we are left with just three independent RG variables which
are $\kappa_{1}^{(m)}$, $\tau_{0}^{(m)}$ and
$\lambda_{1}^{(m)}$ governed by the RG equations
\begin{eqnarray}
   \kappa_{1}^{(m+1)} \>=\> \kappa_{1}^{(m)} + \kappa_{m+2} - \kappa_{m+1} - 2(\lambda_{1}^{(m)} - 
   \lambda_{m+1}) - \frac{2[\tau_{0}^{(m)}]^{2}}{\kappa_{1}^{(m)}+ \tau_{m+1}} \quad 
   & \forall m \in \{0,\ldots,k-2\}  \label{kitten}\\
   \tau_{0}^{(m+1)} \>=\> \lambda_{1}^{(m)} - 
   \frac{[\tau_{0}^{(m)}]^{2}}{\kappa_{1}^{(m)}+ \tau_{m+1}} 
   \quad & \forall m \in \{0,\ldots,k-1\} \label{calf}\\
   \lambda_{1}^{(m+1)} \>=\> \lambda_{m+2} + 
   \frac{[\tau_{0}^{(m)}]^{2}\tau_{m+1}}{[\kappa_{1}^{(m)}]^{2} 
   - \tau_{m+1}^{2}} \quad & \forall m \in \{0,\ldots,k-1\} \label{puppy}
.
\end{eqnarray}
\end{widetext}

\section{Analysis of the RG Equations \label{analysis}}

\subsection{One-dimensional Lattice (HN2)\label{sub:HN2trans}}
It will prove helpful to demonstrate the general set of
recursions [Eqs.~(\ref{kitten}, \ref{calf}, \ref{puppy})] by way of the one-dimensional ($d=1$) 
ring of $N=2^{k}$ sites.  For consistency we call this the HN2 network, or Hanoi network 
with 2 bonds per site (with additional bonds for the input/output leads in the 
ring geometry).  
We can employ the recursions to explore the transmission through a $d=1$ ring of
$N=2^k$ sites.  The energy scale chosen throughout is such that the 
(uniform) transmissivity for each bond has a unit weight.  
With that, we obtain the initial conditions 
\begin{eqnarray}
\kappa_{i}^{(0)} & = & E,\qquad(i\geq1),\nonumber \\
\tau_{0}^{(0)} & = & -1,\label{eq:transIC_HN2}\\
\tau_{i}^{(0)} & = & 0,\qquad(i\geq1),\nonumber \\
\lambda_{i}^{(0)} & = & 0,\qquad(i\geq1)\>.\nonumber
\end{eqnarray}
Eqs.~(\ref{kitten}, \ref{calf}, \ref{puppy}) simplify to
\begin{eqnarray}
\kappa_{m+1} & = &
\kappa_{m}-\frac{2\tau_{m}^{2}}{\kappa_{m}},
\label{eq:RG-HN2}\\
\tau_{m+1} & = & -\frac{\tau_{m}^{2}}{\kappa_{m}},
\nonumber
\end{eqnarray}
where $\kappa_m\equiv \kappa_{i}^{(m)}$ and
$\tau_{i}=\lambda_{i}\equiv0$ for all $i\geq1$ and
$\tau_m\equiv \tau_{0}^{(m)}$. These nonlinear recursions are
easily solved by defining $s_{m}=-\kappa_{m}/\tau_{m}$ for
which $s_{m+1}=s_{m}^{2}-2$, obtained by dividing the
$2^{\rm nd}$ by the $3^{\rm rd}$ line in Eqs.~(\ref{eq:RG-HN2}).
Formally, the solution is
\begin{equation}
s_{m}=2\cos\left[2^{m}\arccos\left(\frac{\kappa^{(0)}}{2\tau^{(0)}}\right)\right]=
2T_{2^{m}}\left(\frac{\kappa^{(0)}}{2\tau^{(0)}}\right),
\label{eq:sn}
\end{equation}
where $T_{n}(x)$ refers to the $n$-th Chebyshev polynomial of
the first kind \citep{abramowitz:64}. Inserting into
Eqs.~(\ref{eq:RG-HN2}) and applying the initial conditions in
Eqs.~(\ref{eq:transIC_HN2}), generates the results
\begin{eqnarray}
\tau^{(m)} & = & -\prod_{i=0}^{m-1}\frac{1}{s_{i}},\nonumber \\
\kappa^{(m)} & = & s_{m}\prod_{i=0}^{m-1}\frac{1}{s_{i}},
\label{eq:HN2solutions}
\end{eqnarray}
where the last equality emerges under reordering factors in the
products.

Eq.~(\ref{linear1tT}) in Appendix~\ref{Sec:AppB} shows that the
transmission amplitude $t$ is directly proportional to 
$\tau^{(k)}$.  Clearly, if there is no transmission on any bond, 
{\it i.e.\/} $\tau^{(k)}=0$, for a given input energy $E$, there can be no
transmission through the network itself, no matter what happens on the
sites.  But
instead of plotting $\tau^{(k)}$, it will prove more instructive to 
plot $\kappa^{(k)}$.  It is easy to see from 
Eqs.~(\ref{eq:HN2solutions}) that $\kappa^{(k)}$ varies rapidly 
whenever $\tau^{(k)}$ does, but that $\kappa^{(k)}$ varies smoothly 
whenever $\tau^{(k)}$ vanishes.  In the following, we will see that
this behavior remains true for HN3 and HN5, in which case the
variation of $\kappa^{(k)}$ with $\kappa^{(0)}=E$ provides more 
information beyond the mere vanishing of $\tau^{(k)}$.  

\begin{figure}
\includegraphics[bb=0bp 0bp 290bp 185bp,clip,scale=0.8]{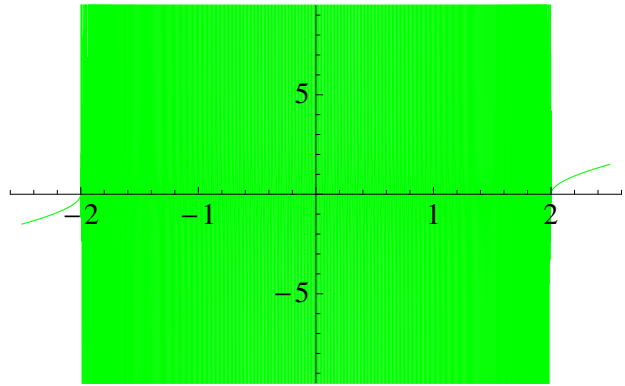}
\caption{\label{fig:q10HN2} (Color online) Plot of $\kappa^{(10)}$ in 
Eq.~(\ref{eq:RG-HN2}) as a function of its initial condition 
$\kappa^{(0)}=E$. 
Even at this small order, the function varies extremely rapidly, 
such that its (green-shaded) line completely covers the shown domain.  
Thus, for $-2\leq E\leq2$,
$\kappa^{(m)}$ for any sufficiently large $m$ is a random function.
Correspondingly, the transmission spectrum is dense, with full transmission
close to any input energy $E$ such that particles do not localize.}
\end{figure}

We have evolved the RG-recursion in~(\ref{eq:RG-HN2}) for the
initial conditions in~(\ref{eq:transIC_HN2}) and plotted
$\kappa^{(k=10)}$ as a function of $\kappa^{(0)}=E$ in
Fig.~\ref{fig:q10HN2}. Even at that system size,
$N=2^{k}=1024$, 
delocalized states completely cover the domain $-2\leq E\leq2$.

\subsection{Case HN3\label{sub:Case-HN3transm}}
\begin{figure}
\includegraphics[bb=0bp 0bp 290bp 193bp,clip,scale=0.8]{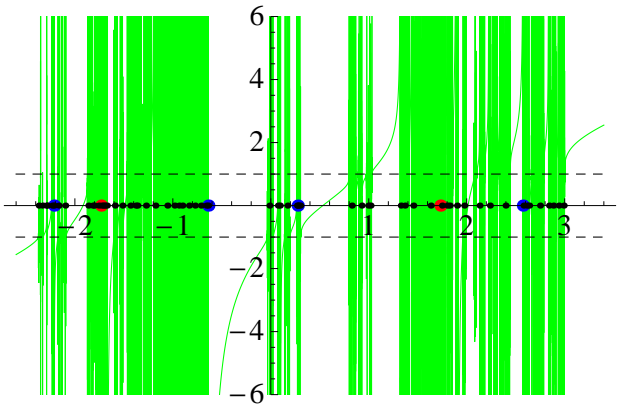}
\caption{\label{fig:q200HN3}
(Color online) Plot of $\kappa_{200}$ in 
Eq.~(\ref{eq:transmRG-HN3_redux}) as a function of its
initial condition $\kappa_{0}=E$. 
Bands of localized and delocalized
states intermix. Correspondingly, there are
localization-delocalization transitions already before the addition of
any additional randomness in HN3, merely as a function of the input energy $E$. On
the horizontal axis, we have marked the solutions $E_{s}^{(i)}$ of
Eq.~(\ref{eq:qcritical}) for $s=1$ (red dots), $s=2$ (blue dots), and
$s=6$ (small black dots). The accumulation of the latter demonstrates
(even for such a small value of $s$) that the band gaps are associated
with the absence of such solutions. While the solutions for $s=1$
happen to be interior to the bands, some of those for $s=2$ appear to
mark the band edges, in particular the one at
$E=E_{2}^{(2)}=-0.637875$.}
\end{figure}

Again, we can employ the RG Eqs.~(\ref{kitten}, \ref{calf}, \ref{puppy}) for
transmission through HN3 consisting of a ring of $N=2^{k}$
sites, as in Fig.~\ref{fig:3hanoi}. Since all initial diagonal entries 
are identical, the hierarchy for the $\kappa_{i}$ collapses and we retain only 
two nontrivial relations, one for $\kappa_{1}$ and one for all
other $\kappa_{i}\equiv \kappa_{2}$ for all $i\geq2$. Here, all
$\tau_{i}$ are non-zero, encompassing the backbone links
($i=0$) and all levels of long-range links ($i\geq1$). But it
remains $\tau_{i}\equiv-1$ for $i\geq1$ at any step $m$ of the
RG, in particular, $\tau_{1}^{(m)}\equiv-1$ throughout; only
the backbone $\tau_{0}$ renormalizes non-trivially. Although
all links of type $\lambda_{i}$ are initially absent in this
network, the details of the RG calculation shows that under
renormalization terms of type $\lambda_{1}$ emerge while those
for $\lambda_{i}$ for $i\geq2$ remain zero at any step. Thus,
we obtain far more elaborate RG recursion equations compared to
those of HN2. Abbreviating $\kappa_m\equiv \kappa_{1}^{(m)}$,
$\tau_m\equiv \tau_{0}^{(m)}$, and
$\lambda_m=\lambda_{1}^{(m)}$, 
Eqs.~(\ref{kitten}, \ref{calf}, \ref{puppy}) and their
initial conditions reduce to
\begin{equation}
\begin{matrix}
\kappa_{m+1} & = &
\kappa_{m}-2\lambda_{m}-\frac{2\tau_{m}^{2}}{\kappa_{m}-1}, 
                                        & \qquad & (\kappa_{0}=E),  \\
\tau_{m+1} & = &
\lambda_{m}-\frac{\tau_{m}^{2}}{\kappa_{m}-1},              
                                        & \qquad & (\tau_{0}=-1),   \\
\lambda_{m+1} & = &-\frac{\tau_{m}^{2}}{\kappa_{m}^{2}-1},  
                                        & \qquad & (\lambda_{0}=0). \\
\end{matrix}
\label{eq:transmRG-HN3_redux}
\end{equation}
We have evolved the RG-recursion in~(\ref{eq:transmRG-HN3_redux}) 
and plotted $\kappa_{k=200}$ as a
function of $\kappa_{0}=E$ in Fig.~\ref{fig:q200HN3}. Even at
that enormous (and definitely asymptotic) system size,
$N=2^{200}\approx10^{70}$, domains of localized states remain
asymptotically inside the physically relevant domain of $-2\leq
E\leq2$.  For comparison, the radius of the visible 
universe is only about $10^{40}$~fm.  

In the next subsection, we will explore the asymptotic properties
of these recursions for large $m$. We will find domains in $E$
of stationary solutions, which are particular for HN3, and show
that the special points where this analysis fails correspond to
transitions between localized and delocalized behavior.

\subsubsection{Analysis of the Steady State:\label{sub:SteadyStateAnalysis}}
We can analyze the absorbing steady state, which is the unique
feature of HN3 (in contrast to HN2 and HN5) leading to band
gaps, as follows. Numerical trails show that the RG recursions
in Eq.~(\ref{eq:transmRG-HN3_redux}) reach a steady state for
certain initial conditions $E$ when for all $m$ larger than
some $m_{0}$ it is
\begin{eqnarray}
1 & \gg &
\lambda_{m}\gg\frac{\tau_{m}^{2}}{\kappa_{m}-1},\qquad(m\to\infty).
\label{eq:Asymp}
\end{eqnarray}
The leading contribution for $\kappa_{m}$ in
Eq.~(\ref{eq:transmRG-HN3_redux}) then suggests
\begin{eqnarray}
\kappa_{m+1} & \sim & \kappa_{m}\sim \kappa_{\infty},
\label{eq:qinfty}
\end{eqnarray}
which is a constant that is difficult to derive from the
initial conditions, $\kappa_{0}=E$, unfortunately.

From the recursion for $\tau_{m}$ in
Eq.~(\ref{eq:transmRG-HN3_redux}), we further obtain
\begin{eqnarray}
\tau_{m+1} & \sim & \lambda_{m}  \sim  -\frac{\tau_{m-1}^{2}}{\kappa_{\infty}^{2}-1}.
\label{eq:pasymp}
\end{eqnarray}
This 2nd order difference equation has two solutions, of which
we discard the oscillatory one, to get for large $m>m_{0}$:
\begin{eqnarray*}
\tau_{m} & \sim & -\left(\kappa_{\infty}^{2}-1\right)\exp\left\{ -C\sqrt{2^{m}}\right\} ,\\
\lambda_{m} & \sim & -\left(\kappa_{\infty}^{2}-1\right)\exp\left\{
-C\sqrt{2^{m+1}}\right\} ,
\end{eqnarray*}
where $C>0$ is another undetermined constant that depends on
$E$. Using the recursion for $\kappa_{m}$ in
Eq.~(\ref{eq:transmRG-HN3_redux}) to next-to-leading order
yields
\begin{eqnarray}
\kappa_{m+1} & \sim & \kappa_{m}-2\lambda_{m},
\label{eq:q_correction}\\
 & \sim & \kappa_{m}+2\left(\kappa_{\infty}^{2}-1\right)\exp\left\{
-C\sqrt{2^{m+1}}\right\} ,
\end{eqnarray}
which, when summed from a $m>m_{0}$ to $\infty$ results in
\begin{eqnarray}
\kappa_{m} & \sim &
\kappa_{\infty}-2\left(\kappa_{\infty}^{2}-1\right)\exp\left\{
-C\sqrt{2^{m+1}}\right\} ,
\label{eq:q_asymp}
\end{eqnarray}
where we have kept only the first term in the sum, as the
summand is exponentially decaying.

The consequences of this analysis are quite dramatic.  If such a
steady-state solution is reached, both transmission rates
$\tau$ and $\lambda$ vanish, only leaving a finite on-site
energy $\kappa_{\infty}$. Hence, there can not be any
transmission through the network when such a state is reached,
and gaps emerge in the transmission spectrum. Since the system
size is given by $N=2^{m}$, this result implies that finite
size corrections scale with $\exp\left\{ -C\sqrt{2N}\right\} $,
{\it i.e.\/} finite-size corrections decay rapidly with a stretched
exponential. Note that there are no steady-state solutions that
cross $\kappa_{\infty}=\pm1$ (dashed lines in
Fig.~\ref{fig:q200HN3}), where the correction in
Eq.~(\ref{eq:q_asymp}) would break down. Instead, the approach
of $\kappa_{\infty}\to\pm1$ frequently appears to be associated
with the emergence of a band edge between localized and
delocalized states.

\subsubsection{Band-Edge Analysis:\label{sub:BandEdgeAnalysis}}
The non-trivial band structure warrants some further
investigation. In particular, we can associate such band edges
with initial conditions $E=E_{s}^{(i)}$ for which there exists
a $m=s$ such that
\begin{eqnarray}
\kappa_{s}\left(E_{s}^{(i)}\right) & = & -1,
\label{eq:qcritical}
\end{eqnarray}
a singular point in the recursion for $\lambda_{s+1}$.
{[}Interestingly, any singularity at $\kappa_{m}=+1$ appears to
be benign in that it does \emph{not} affect the continuity in
$\kappa_{m}$ as a function of $E$ for $m\to\infty$; it afflicts
each quantity in Eqs.~(\ref{eq:transmRG-HN3_redux})
\emph{simultaneously}, leading to a divergence in
$\kappa_{m+1}$, $\tau_{m+1}$, and $\lambda_{m+1}$ just so that
$\kappa_{m+2}\approx \kappa_{m-1}$, $\tau_{m+2}\approx
\tau_{m-1}$, $\lambda_{m+2}\approx \lambda_{m-1}$.{]} In
Fig.~\ref{fig:q200HN3}, we have also marked the real solutions
$E_{s}^{(i)}$ of Eq.~(\ref{eq:qcritical}) for $s=1$, 2, and 6.
Clearly, those solutions strongly correlate with the bands, and
there appear to be none within the gaps (although we have not been 
able to prove this conjecture). But while those solutions for $s=1$,
$E_{1}^{(1,2)}=\pm\sqrt{3}$, are located well within some band
(as are those for $s=6$), the four real solutions of
Eq.~(\ref{eq:qcritical}) for $s=2$ satisfy the quartic equation
\begin{eqnarray}
0 & = & 1-2E-6E^{2}+E^{4}
\label{eq:quarticQ}
\end{eqnarray}
and appear to be all associated with some more or less significant
band edge, see blue dots in Fig.~\ref{fig:q200HN3}. We can
speculate that there is a whole hierarchy of transitions, each
associated with one of the solutions $E_{m}^{(i)}$, which may
become dense on certain intervals. While we don't know what
determines those intervals precisely, we can analyze the
behavior in the neighborhood of Eq.~(\ref{eq:qcritical}).  We
observed that the recursions in
Eq.~(\ref{eq:transmRG-HN3_redux}) possess stable steady-state
solutions for large $m$ characterized by $\tau_{m}\sim
\lambda_{m}\to0$, {\it i.e.\/} vanishing bond-strength between input
and output. These solutions prevail in the observed band gaps,
which accordingly correspond to localized states. It seems that
the reason for the persistence of gaps derives from that
stability: band gaps emerge whenever the steady state is
reached \emph{before} Eq.~(\ref{eq:qcritical}) can be
satisfied.  For instance, in the case of HN5 below, any
putative steady-state solution proves unstable for sufficiently
large $m$ such that any band gaps are transitory only, see
Fig.~\ref{fig:q10HN5}.

For the analysis of the recursion
Eq.~(\ref{eq:transmRG-HN3_redux}), we assume that for some
$m=s$, we reach
\begin{eqnarray}
\kappa_{s} & \sim & -1+\epsilon,\qquad(\epsilon\ll1),
\label{eq:qAnsatz}
\end{eqnarray}
where $\epsilon=\epsilon(E)$ may be of either sign, depending
on $\Delta E=E-E_{s}^{(i)}$. Generically,
$\epsilon\propto\Delta E$, see Sec.~\ref{sub:Scaling-Relation:}
below. Assuming that $\tau_{s},\lambda_{s}\ll1/\epsilon$ leads
to
\begin{eqnarray}
\kappa_{s+1} & \sim & -1-2\lambda_{s}+\tau_{s}^{2}+O(\epsilon),\nonumber \\
\tau_{s+1} & \sim & \lambda_{s}+\frac{\tau_{s}^{2}}{2}+O(\epsilon),
\label{eq:s1}\\
\lambda_{s+1} & \sim & \frac{\tau_{s}^{2}}{2\epsilon}+\frac{\tau_{s}^{2}}{4}+O(\epsilon),
\nonumber
\end{eqnarray}
which leaves only $\lambda_{s+1}$ singular. After one more
recursion step, we get instead
\begin{eqnarray}
\kappa_{s+2} & \sim &
-\frac{\tau_{s}^{2}}{\epsilon}+\frac{2+6\lambda_{s}+2\lambda_{s}^{2}-2\tau_{s}^{2}-5\lambda_{s}
\tau_{s}^{2}}{-2-2\lambda_{s}+\tau_{s}^{2}}+O(\epsilon),
\nonumber \\
\tau_{s+2} & \sim &
\frac{\tau_{s}^{2}}{2\epsilon}+
\frac{2\lambda_{s}^{2}+\tau_{s}^{2}+3\lambda_{s}\tau_{s}^{2}}
     {4+4\lambda_{s}-2\tau_{s}^{2}}+O(\epsilon),
\label{eq:s2}\\
\lambda_{s+2} & \sim & O(1).
\nonumber
\end{eqnarray}
At this point, 
Eq.~(\ref{eq:transmRG-HN3_redux}) decouple to
leading order, as $\lambda_{s+2+i+1}\sim
-\tau_{s+2+i}^{2}/\kappa_{s+2+i}^{2}\sim -1/4$ remains of order
$O(1)$ while both $\kappa_{s+2+i}$ and $\tau_{s+2+i}$ are of
order $O(1/\epsilon)$, and we get for some $i\geq0$
\begin{eqnarray}
\kappa_{s+2+i+1} & \sim & \kappa_{s+2+i}-\frac{2\tau_{s+2+i}^{2}}{\kappa_{s+2+i}},
\nonumber \\
\tau_{s+2+i+1} & \sim & -\frac{\tau_{s+2+i}^{2}}{\kappa_{s+2+i}}.
\label{eq:asymp_qp}
\end{eqnarray}
These are \emph{exactly} the same recursions we obtained in
Eq.~(\ref{eq:RG-HN2}) for HN2, with the solution in
Eq.~(\ref{eq:sn}):
\begin{eqnarray}
\frac{\kappa_{s+2+i}}{2\tau_{s+2+i}} & \sim & 
   -T_{2^{i}}\left(-\frac{\kappa_{s+2}}{2\tau_{s+2}}\right),
\label{eq:asymp_T}
\end{eqnarray}
where from Eq.~(\ref{eq:s2}) we have
\begin{eqnarray}
\label{eq:asymp_IC}
\frac{\kappa_{s+2}}{2\tau_{s+2}} & \sim &
-1+A\frac{\epsilon}{\tau_{s}^{2}}, \\
A & = &  \frac{2+6\lambda_{s}
-3\tau_{s}^{2}-8\lambda_{s}\tau_{s}^{2}}{2+2\lambda_{s}-\tau_{s}^{2}}
\>.
\end{eqnarray}
With that inserted into Eq.~(\ref{eq:asymp_T}), we can deduce
\begin{eqnarray}
\frac{\kappa_{s+2+i}}{2\tau_{s+2+i}} & \sim &-T_{2^{i}}\left(1-A\frac{\epsilon}{\tau_{s}^{2}}\right),
\nonumber \\
 & \sim & 1-A\frac{\epsilon}{\tau_{s}^{2}}T_{2^{I}}'\left(1\right),
\nonumber \\
 & \sim & 1-2^{2i}A\frac{\epsilon}{\tau_{s}^{2}},
\label{eq:asymp_correction}
\end{eqnarray}
since $T'_{n}(x)=nU_{n-1}(x)$ and $U_{n-1}(1)=n$, referring to
the Chebyshev polynomial of the 2nd kind, $U_{n}(x)$
\citep{abramowitz:64}. With the exponential growth in $i$ of
the correction amplitude in the asymptotic expansion in
Eq.~(\ref{eq:asymp_correction}), the expansion breaks down at
some $i\sim i_{0}$ such that the correction itself becomes of
$O(1)$, {\it i.e.\/} 
\begin{eqnarray}
i_{0} & \sim & \frac{1}{2}\log_{2}\left(\frac{\tau_{s}^{2}}{\left|A\epsilon\right|}\right).
\label{eq:i0}
\end{eqnarray}
For $i>i_{0}$, according to the first line of
Eq.~(\ref{eq:asymp_correction}) the ratio 
$\kappa_{s+2+i}/\tau_{s+2+i}$ either rises or falls
exponentially, depending on whether $A\epsilon<0$ or
$A\epsilon>0$, respectively.  In the latter case, $\kappa$
becomes less relevant and the bonds $\tau$ and $\lambda$
determine the future evolution in $m$, leading again to the
chaotic behavior in $\kappa_{m}$ observed within the bands in
Fig.~\ref{fig:q200HN3}. On the other hand, if $A\epsilon>0$,
the on-site energies $\kappa$ dominate exponentially over the
couplings $\tau$ and $\lambda$, evolving towards an absorbing
steady state on the band-gap side of the transition.

\subsubsection{Scaling Relation for $\kappa_{\infty}(E)$:\label{sub:Scaling-Relation:}}
As mentioned in Sec.~\ref{sub:SteadyStateAnalysis}, we can not
generally predict the dependence of the asymptotic behavior on
the initial condition $E$. But we can use the analysis of that
section at least to determine the behavior of
$\kappa_{\infty}(E)$ on the approach to those band edges where
it diverges.

\begin{figure}
\includegraphics[bb=0bp 0bp 300bp 190bp,clip,scale=0.7]{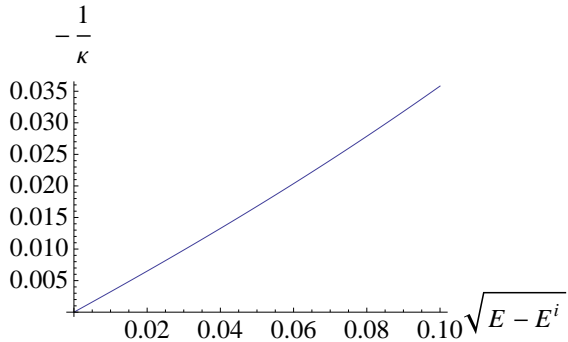}
\caption{\label{fig:qinftyDQ} (Color online) 
Plot of $1/\kappa_{\infty}$ as a function of $\sqrt{E-E_{s}^{(i)}}$
for $E\to E_{s}^{(i)}$ to test Eq.~(\ref{eq:qinftyDQ}). Here,
$E_{2}^{(2)}=-0.637875$, marked as the 2nd blue dot from the left in
Fig.~\ref{fig:q200HN3}.}
\end{figure}

It is easy to show, using just the singular terms for $\kappa$
and $\tau$ in Eq.~(\ref{eq:s2}) inserted into the recursions
in Eq.~(\ref{eq:asymp_qp}), that both simultaneously decay
exponentially with $i$ at least while $i\lesssim i_{0}$ from
Eq.~(\ref{eq:i0}),
\begin{eqnarray}
\kappa_{s+2+i} & \sim & -\frac{\tau_{s}^{2}}{2^{i}\epsilon}+O(1),
\label{eq:qs_asymp}\\
\tau_{s+2+i} & \sim & \frac{\tau_{s}^{2}}{2^{i+1}\epsilon}+O(1).
\nonumber
\end{eqnarray}
At $i\sim i_{0}$ there is a cross-over beyond which for all
$i>i_{0}$ it is $\tau_{s+2+i}\to0$ and a saturated value of
$\kappa_{s+2+i}\to \kappa_{\infty}$ is reached. We can obtain
the dominant asymptotic behavior at the cross-over from
\begin{eqnarray}
\kappa_{\infty} & \sim & \kappa_{s+2+i_{0}} \sim 
-\frac{\tau_{s}^{2}}{\epsilon\sqrt{\frac{\tau_{s}^{2}}{\left|A\epsilon\right|}}},
\nonumber \\
 & \sim &
-\sqrt{\left|A\right|\tau_{s}^{2}}\frac{\mathrm{sgn}\left(\epsilon\right)}
{\left|\epsilon\right|^{\frac{1}{2}}}.
\label{eq:qinfty_eps}
\end{eqnarray}
We can further establish a (generic) relation between
$\epsilon$ and
\begin{eqnarray*}
\Delta E & \sim & E-E_{s}^{(i)}
\end{eqnarray*}
by extending the discussion of Eq.~(\ref{eq:qcritical}) in
Sec.~\ref{sub:BandEdgeAnalysis}. We set
\begin{eqnarray*}
-1+\epsilon\left(\Delta E\right) & \sim & \kappa_{s}\left(E_{s}^{(i)}+\Delta E\right),\\
 & \sim & \kappa_{s}\left(E_{s}^{(i)}\right)+\kappa'_{s}\left(E_{s}^{(i)}\right)\Delta E,\\
 & \sim & -1+\kappa'_{s}\left(E_{s}^{(i)}\right)\Delta E,
\end{eqnarray*}
since $\kappa_{s}\to-1$ is a regular limit for $\Delta E\to0$,
as Eq.~(\ref{eq:quarticQ}), for example, suggests. Hence,
\begin{eqnarray*}
\epsilon\left(\Delta E\right) & \sim &
\kappa'_{s}\left(E_{s}^{(i)}\right)\Delta E,
\end{eqnarray*}
and we conclude
\begin{eqnarray}
\kappa_{\infty} & \sim &
-\sqrt{\frac{\left|A\right|\tau_{s}^{2}}
{\left|\kappa'_{s}\left(E_{s}^{(i)}\right)\right|}}
\frac{\mathrm{sgn}\left(\kappa'_{s}\left(E_{s}^{(i)}\right)\Delta
  E\right)}{\left|\Delta E\right|^{\frac{1}{2}}}
\label{eq:qinftyDQ}
\end{eqnarray}
with $\Delta E\to0$.  
In Fig.~\ref{fig:qinftyDQ} we have tested the asymptotic
relation $1/\kappa_{\infty}\sim\sqrt{\left|\Delta E\right|}$ in
the band gap near $E_{s}^{(2)}=-0.637875\ldots$, a solution of
Eq.~(\ref{eq:quarticQ}) marked blue in Fig.~\ref{fig:q200HN3}.

\subsection{Interpolation between HN2 and HN3\label{sub:Interpolation-between-HN2}}
It proves fruitful to consider an interpolation between the
case of HN2 in Sec.~\ref{sub:HN2trans} and HN3 in
Sec.~\ref{sub:Case-HN3transm} in terms of a one-parameter
family of models. To wit, we can accomplish such an
interpolation by weighting the transmission along the
small-world links (see red-shaded links in
Fig.~\ref{fig:HN3scatterKoch}) by a factor of $y$ relative to
that of the backbone links (see black links in
Fig.~\ref{fig:HN3scatterKoch}). Clearly, more generally,
hierarchy and/or distance-dependent weights could be introduced
as well. For $y=0$, small-world links are non-existent, and we have 
the linear lattice HN2. Although still mostly delocalized, the states of the
systems immediately change behavior when $y>0$, and we find
localized states which expand their domain towards $y=1$,
corresponding to HN3, and continue to do so until at about
$y=4$ no transmission is possible any longer: The more we
weight small-world links here, which classically would
\emph{expedite} transport \citep{TASEP09}, the less quantum
transport is possible! In the next section we will see that
even more small-world links, as in HN5, can lead to more
transmission again. Hence, the detailed structure of the links
matter.

To explore this $y$-family of models, we have to generalize
Eq.~(\ref{eq:transmRG-HN3_redux}) appropriately:
\begin{equation}
\label{eq:RG-HN2y}
\begin{matrix}
\kappa_{m+1}  & = & 
               \kappa_{m}-2\lambda_{m}-\frac{2\tau_{m}^{2}}{\kappa_{m}-y}, 
                & \qquad & (\kappa_{0}=E), \\
\tau_{m+1}    & = &
               \lambda_{m}-\frac{\tau_{m}^{2}}{\kappa_{m}-y}, 
                & \qquad & (\tau_{0}=-1), \\
\lambda_{m+1} & = & -\frac{y\,\tau_{m}^{2}}{\kappa_{m}^{2}-y^{2}}, 
                & \qquad & (\lambda_{0}=0), 
\end{matrix}
\end{equation}
since for all non-backbone links in Eqs.~(\ref{kitten}, \ref{calf}, \ref{puppy})
it is $\tau_{i\geq1}^{(0)}=y\, \tau_{0}^{(0)}$, {\it i.e.\/} 
$\tau_{i\geq1}^{(m)}=-y$ at every RG-step. Note that these
equations reduce to Eqs.~(\ref{eq:RG-HN2}) for $y=0$ (with all
$\lambda_{i}\equiv0$) and to Eq.~(\ref{eq:transmRG-HN3_redux})
for $y=1$.

\begin{figure*}
\includegraphics[bb=0bp 0bp 590bp 550bp,clip,scale=1.0]{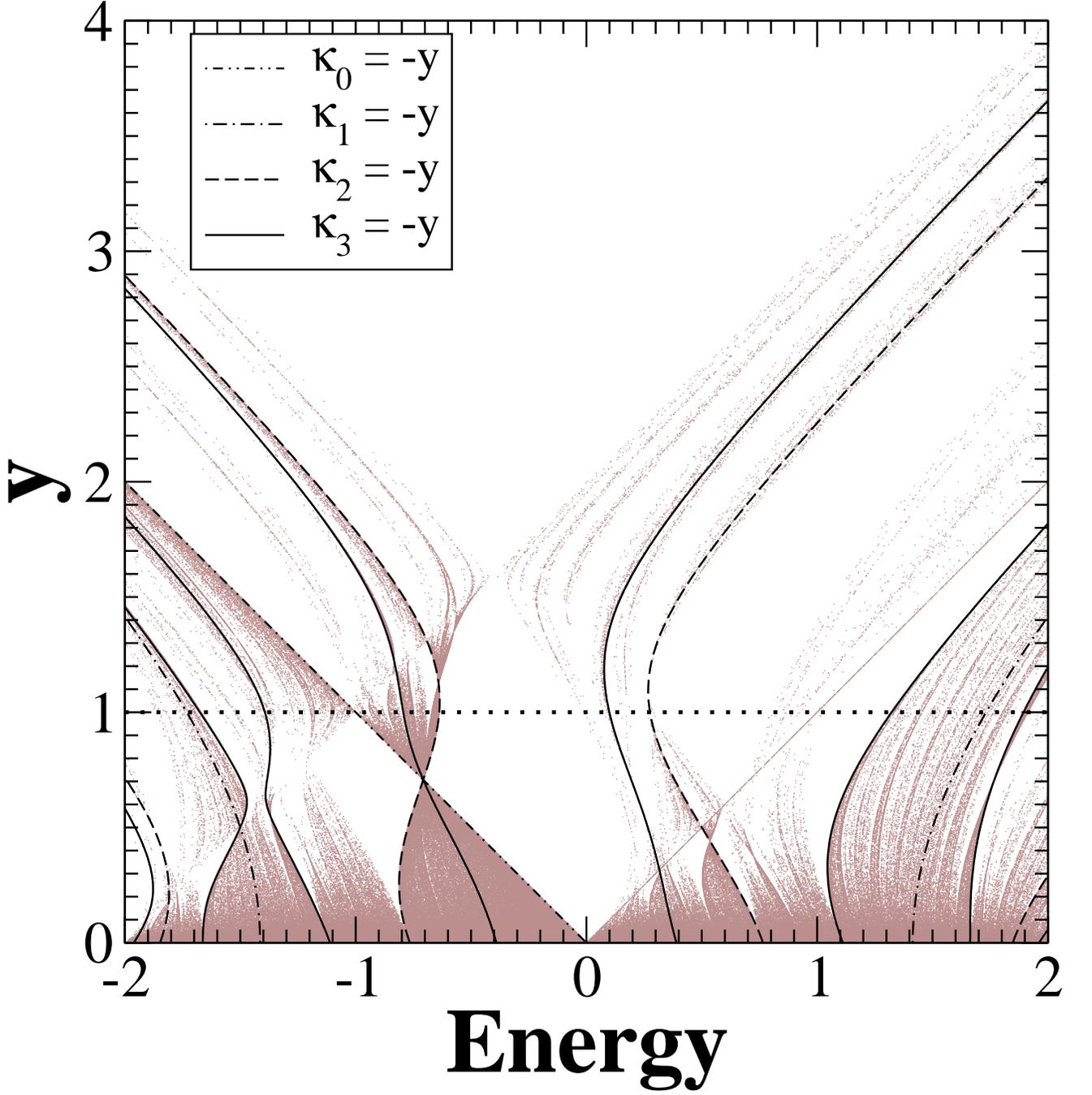}
\caption{\label{fig:HN2yplot} (Color online) Plot of the state of
Eq.~(\ref{eq:RG-HN2y}) after 1000 iterations for initial
energies $\kappa_{0}=E$ and the interpolation parameter $y$
(the resolution is 0.001 in each direction). Shaded points have
not, or not yet, converged to a steady-state, {\it i.e.\/} those
values possess non-zero transmission. At $y=0$, corresponding
to HN2, the system transmits for any input energy, $-2\leq
E\leq2$. As soon as $y>0$, bands of localized states emerge
(especially at $E=0$), and the remaining transmitting states
exhibit a chaotic dependence on the parameters. At $y=1$, the
case of HN3 marked by a dotted horizontal line, only a few
non-localized states remain, and the further strengthening of
small-world links diminish transmission even more, such that it
ceases completely for $y>4$. Any band of transmitting states
appears to be accompanied by solutions of
Eq.~(\ref{eq:y_qcritical}), which here are draw as curves for
the lowest orders of the recursion only; $s=0,1,2,$ and~3 
corresponding respectively to dot-dot-dashed, dot-dashed, dashed, 
and solid curves.}
\end{figure*}

In Fig.~\ref{fig:HN2yplot} we map out the state of
Eq.~(\ref{eq:RG-HN2y}) after the $1000^{\rm th}$ iteration based
on whether a steady state has been reached or not, depending on
the incoming energy $E$ and the relative weight $y$. For any
$y>0$, the ability to transmit has a strong chaotic dependence
on these parameters, and ceases completely for $y>4$. Even
within domains of apparent transmission there are often
sub-domains where no transmission is possible, and it is not
clear whether true conduction bands exist. Since in this model
the long-range links are not connected to each other except
through the backbone, one may speculate that even at high weight
these links merely lead to localized resonances that interfere
with transport along the backbone instead of conveying it. (A
similar confinement effect was observed for the RG applied to
random walks on HN3 in Ref.~\citep{SWN}.)

It is straightforward to generalize the discussion for HN3 in
Sec. \ref{sub:Case-HN3transm} to this model. In particular, for
the band-edge analysis we have to generalize
Eq.~(\ref{eq:qcritical}) to read
\begin{eqnarray}
\kappa_{s}\left(E_{s}^{(i)}\right) & = & -y.
\label{eq:y_qcritical}
\end{eqnarray}
For $s=0,1,2$, and~3, the numerical solutions $E_{s}^{(i)}(y)$
are also plotted as lines in Fig.~\ref{fig:HN2yplot}. The
result underlines the contention made before for HN3 that the
transitions between transmission and localization are closely
associated with these singular points of the
Eq.~(\ref{eq:RG-HN2y}). For instance, the big blue-shaded dots
in Fig.~\ref{fig:q200HN3} correspond here to the intersection
of the simple-dashed line for $s=2$ with the dotted horizontal
line along $y=1$ (i.~e. HN3). Dominant features emerge, such as
the line $\kappa_{0}=E=-y$. Other interesting points become
apparent, for instance, the one at $y=-E=1/\sqrt{2}$. While
there is otherwise no apparent relation to solutions of
$\kappa_{s}=+y$, it should be noted that its $s=0$ case
\emph{does} produce a distinct feature in the line $E=y$.

\subsection{Case HN5\label{sub:Case-HN5transm}}
In close correspondence with the treatment in
Sec.~\ref{sub:Case-HN3transm}, we can employ the RG in
Eqs.~(\ref{kitten}, \ref{calf}, \ref{puppy}) for transmission through HN5
consisting of a ring of $N=2^{k}$ sites. The sole difference
with Sec.~\ref{sub:Case-HN3transm} is that all links of type
$\lambda_{i}$ are initially present in this network. Yet, the
details of the RG calculation in Sec.~\ref{MatrixRG} show that
under renormalization only links of type $\lambda_{1}$
renormalize while those $\lambda_{i}$ for $i\geq2$ remain
unrenormalized at any step. The diagonal elements are again
hierarchy-independent, $\kappa_{i}^{(0)}\equiv E$, while the
recursion for $\lambda$ changes. Abbreviating $\kappa\equiv \kappa_{1}$, $\tau\equiv
\tau_{0}$, and $\lambda=\lambda_{1}$,
Eqs.~(\ref{kitten}, \ref{calf}, \ref{puppy}) and their
initial conditions reduce to
\begin{equation}
\label{eq:transmRG-HN5_redux}
\begin{matrix}
\kappa_{m+1}  & = &
              \kappa_{m}-2\lambda_{m}-2-2\frac{\tau_{m}^{2}}{\kappa_{m}-1}, 
              & \qquad & (\kappa_{0}=E), \\
\tau_{m+1}    & = & \lambda_{m}-\frac{\tau_{m}^{2}}{\kappa_{m}-1},
              & \qquad & (\tau_{0}=-1), \\
\lambda_{m+1} & = &
              1-\frac{\tau_{m}^{2}}{\kappa_{m}^{2}-1}, 
              & \qquad & (\lambda_{0}=-1).
\end{matrix}
\end{equation}

\begin{figure}
\includegraphics[bb=0bp 0bp 290bp 180bp,clip,scale=0.8]{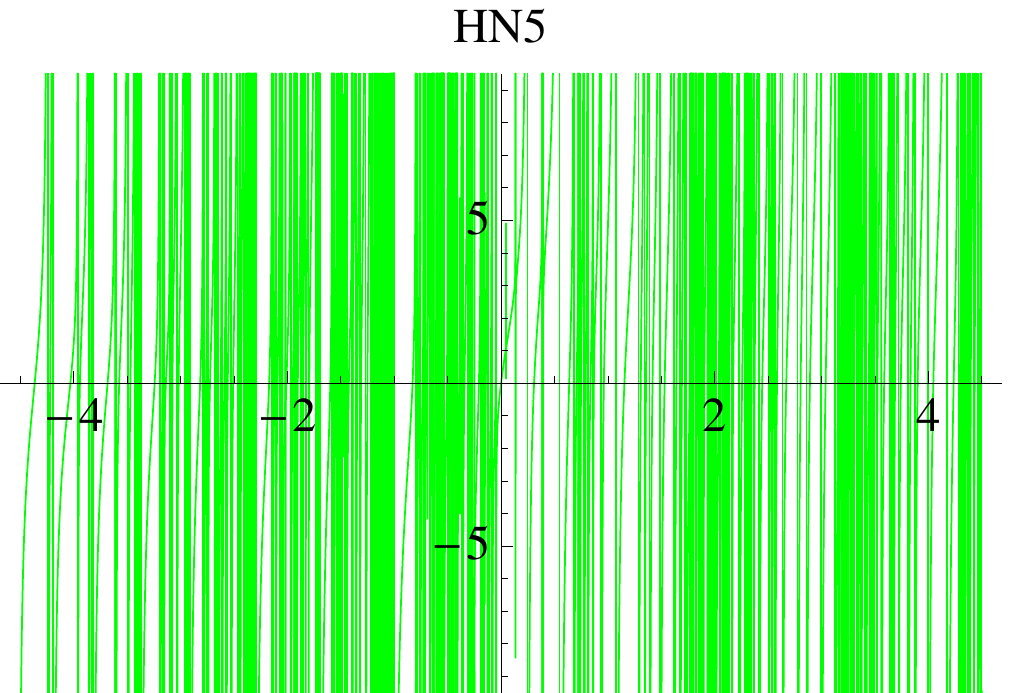}
\caption{\label{fig:q10HN5} (Color online) 
Plot of $\kappa^{(10)}$ in Eqs.~(\ref{eq:transmRG-HN5_redux}) for HN5 
as a function of its initial condition $\kappa^{(0)}=E$. Bands of localized
and delocalized states intermix, but those localized intervals disappear
asymptotically. }
\end{figure}

We have evolved the RG-recursion
in Eq.~(\ref{eq:transmRG-HN5_redux}) and plotted $\kappa^{(k=10)}$
as a function of $\kappa^{(0)}=E$ in Fig.~\ref{fig:q10HN5}.  At
that (definitely not asymptotic) system size,
$N=2^{10}\approx10^{3}$, domains of localized states remain
which will disappear asymptotically, as for the case of HN2 in
Fig.~\ref{fig:q10HN2}. Unlike for HN3, there are no
steady-state solutions for Eq.~(\ref{eq:transmRG-HN5_redux})
that could signal localization. It is interesting to analyze
the cause of this behavior.  Since the $\lambda$-links
[green-shaded in Fig.~\ref{fig:5hanoi}] are an original feature
of the network, the RG recursion
in Eq.~(\ref{eq:transmRG-HN5_redux}) for $\lambda_{m}$ obtain a
constant offset preventing $\lambda_{m}$, and hence $\tau_{m}$,
from vanishing. Unlike for HN3, these small-world links allow
perpetual transmission within a given level of the hierarchy,
instead of interfering with other paths, and transmission is
enhanced.  In fact, we have studied also an interpolation
between HN3 and HN5 by attaching a relative weight $y$ to these
$\lambda$-links in HN5, relative to the otherwise uniform links
present in HN3. Thus, for $y=0$ HN3 is obtained, and for $y=1$
we get HN5. Yet, we find that for any $y>0$, this model
eventually behaves like HN5, with unfettered transmission
throughout the energy spectrum.

\section{Summary and Discussion \label{conclusion}}
We have devised a decimation RG procedure within the matrix methodology of 
Ref.~\citep{daboul} to obtain transmission of quantum electrons through 
networks within the tight-binding model approximation.  
This decimation RG procedure of Appendix~\ref{Sec:AppA} can in principle 
be implemented for any network, in that Eq.~(\ref{Eq:AppA01}) with $n+m$ sites is 
reduced to Eq.~(\ref{Eq:AppA07}) with $n$ sites.  For general 
networks the bookkeeping required could become prohibitive.  However, we find 
the RG equations well suited to the networks we have chosen to analyze, 
namely the hierarchical networks called HN3 (Hanoi Network with 3 bonds per site) 
and HN5 (Hanoi Network with an average of 5 bonds per site).  
This is because the RG can proceed block-wise, as depicted in Fig.~\ref{BuildingBlock}.  
We have analyzed 
both the ring geometry (Fig.~\ref{fig:3hanoi}) and the linear geometry 
(Figs.~\ref{fig:3hanoi} and \ref{fig:HN3scatterKoch}) 
of these networks, with the only difference being the last steps of the RG 
(Appendix~\ref{Sec:AppB}).  We have also analyzed how the transmission for the 
linear lattice (which we label HN2) changes with the strength of the 
small-world-type bonds added to form HN3 (Fig.~\ref{fig:HN2yplot}).  

The Hanoi networks are hierarchical models that provide an intermediary between 
regular lattices and lattices that have random small-world bonds placed on 
regular lattices.  Since the small-world-type bonds in the Hanoi networks 
provide short-cuts between sites, one might expect intuitively that they 
should provide extra paths for transmission.  However, because of the 
hierarchical nature of the networks, the networks no longer possess 
translational symmetry.  This broken symmetry is seen by the incoming 
quantum electrons, and can lead to Anderson-type localization.  
For the HN3 network with variable strength $y$ for the small-world-type bonds, 
we find that the more we weight the small-world links, which classically would 
{\it expedite\/} transport \citep{TASEP09}, the less quantum transport is 
possible (Fig.~\ref{fig:HN2yplot}).  Furthermore, we find that HN3 has 
band edges at particular energies $E$ of the 
incoming electrons, between band gaps with near zero transmission and 
regions of extended wavefunctions and transmission near unity.  
The network HN5 adds still more small-world type bonds to HN3, but we find that 
for any non-zero strength of these additional bonds the band edges seen 
in HN3 disappear and approximately unfettered transmission is seen for 
large enough lattices for any energy of the incoming electrons.  
Thus the hierarchical nature of 
these lattices lead to very interesting transmission properties.  
Thus for these hierarchical lattices, the metal-insulator transitions 
depend on quantities other than just the embedding dimension.  Similar effects 
have been seen for critical phenomena in hierarchical lattices, but only where 
translational symmetry is broken 
\cite{Gefen1980,Banav1983,LeGuillou1987,Novotny1992,Novotny1993}.

Since the HN3 and HN5 networks are planar (see for example 
Fig.~\ref{fig:HN3scatterKoch}), experimental realizations of these 
networks should be possible to construct using etching techniques.  
These experiments would lead to very interesting 
device physics, in particular at the energy-dependent band edges we 
have analyzed for HN3.  

\begin{acknowledgments}
The authors thank Lazarus Solomon for helpful discussions. 
MAN thanks Emory University for hospitality during a one-month 
sabbatical stay.
SB acknowledges support from the U.S.\ National
Science Foundation through grant number DMR-0812204.
\end{acknowledgments}

\appendix

\section{General Matrix Formulation of the RG \label{Sec:AppA}}
As in Ref.~\citep{daboul}, a \lq blob' of atoms in the tight-binding approximation 
is considered to be connected to two semi-infinite leads.  Each semi-infinite 
lead is considered to be a linear arrangement of tight-binding sites with on-site 
energy $0$ and a hopping parameter of $-1$ (setting the energy scale).  
The incoming (outgoing) lead is connected 
to the \lq blob' sites by a vector of hopping parameters ${\vec w}$, 
(respectively, ${\vec u}$).  
The Schr{\"o}dinger equation for the infinite system, 
${\cal H}_\infty {\vec\Psi} = E {\vec\Psi}$ must be solved with the appropriate boundary 
conditions.  
The {\it ansatz\/} made \cite{daboul} is that the wavefunction at every site in the 
in-coming lead has the form $\psi_{m-1}=e^{i m q}+ r e^{-i m q}$ with $m=-\infty,\cdots,-2,-1,0$, 
and the outgoing lead has the wavefunction 
$\psi_{m+1} = t e^{i m q}$ with $m=0,1,2,\cdots,\infty$.  
The wavevector $q$ is related to the energy of the incoming electron by $E=2\cos(q)$.  
The reflection probability is 
$R=|r|^2$ and the transmission probability is $T=|t|^2$.  
With this ansatz, the required solution of the infinite matrix Sch{\"o}dinger equation 
reduces to the solution of a finite matrix equation of dimension two larger than then 
number of sites in the \lq blob'.  
Unlike Ref.~\citep{daboul}, we assume no direct hopping between the leads, {\it i.e.\/} 
no short-cut path around the \lq blob'.  

Consider the case with $n+m$ sites in the blob. We specialize
to the case where all hopping parameters ($\tau$ or $\lambda$) and on-site
energies (convoluted with $E$ to give $\kappa$) are real.  
The $(n+m+2)\times(n+m+2)$ matrix
to solve for the transmission $T=|t|^2$ is \cite{daboul} 
\begin{equation}
\label{Eq:AppA01}
\begin{pmatrix}
\xi           & {\vec w}^{\rm T} & {\vec w}_{d}^{\rm T} & 0             \\
{\vec w}    & {\bf A}            & {\bf B}               & {\vec u}    \\
{\vec w}_{d} & {\bf B}^{\rm T}    & {\bf D}             & {\vec u}_{d} \\
0             & {\vec u}^{\rm T} & {\vec u}_{d}^{\rm T} & \xi           \\
\end{pmatrix}
\begin{pmatrix}
1+r            \\
{\vec \psi}    \\
{\vec \psi}_d  \\
t              \\
\end{pmatrix}
\>=\>
\begin{pmatrix}
2 i \Im (\xi) \\
{\vec 0}_n    \\
{\vec 0}_m    \\
0             \\
\end{pmatrix}
\end{equation}
with $\Im(\xi)$ the imaginary part of the complex function $\xi$, and the definition
\begin{equation}
\label{Eq:AppA02}
\xi = {\rm e}^{iq} - E = -\frac{E}{2} + i \frac{\sqrt{4-E^2}}{2}
.
\end{equation}
The matrix ${\bf A}$ is of size $n\times n$ and includes all
interactions between the $n$ sites that will remain after the
RG. The matrix ${\bf D}$ is of size $m\times m$ and includes
all interactions between the $m$ sites that will be decimated
by the RG. The matrix ${\bf B}$ is of size $n\times m$ and
includes all interactions between the the $n$ sites that will
remain and the $m$ sites that will be decimated. The matrices
${\bf A}$ and ${\bf D}$ are both symmetric matrices, while in
general ${\bf B}$ is not symmetric. The vectors ${\vec w}$,
${\vec u}$, ${\vec \psi}$, and ${\vec 0}_n$ are all of length
$n$, while the vectors ${\vec w}_{d}$, ${\vec u}_{d}$, ${\vec
\psi}_d$, and ${\vec 0}_m$ are all of length $m$.

Multiplying out the two middle rows gives the equations
\begin{equation}
\label{Eq:AppA03}
(1+r){\vec w} +{\bf A}{\vec\psi}+{\bf B}{\vec\psi}_d + t{\vec u} = {\vec 0}_n
\end{equation}
and
\begin{equation}
\label{Eq:AppA04}
(1+r){\vec w}_{d} +{\bf B}^{\rm T}{\vec\psi}+{\bf D}{\vec\psi}_d +
t{\vec u}_{d} = {\vec 0}_m
.
\end{equation}
Solve Eq.~(\ref{Eq:AppA04}) for ${\psi}_d$ to give
\begin{equation}
\label{Eq:AppA05}
{\vec \psi}_d =
-{\bf D}^{-1}
\left[
(1+r){\vec w}_{d} +{\bf B}^{\rm T}{\vec\psi}+ t{\vec u}_{d}
\right]
.
\end{equation}
\begin{widetext}
Substituting ${\vec\psi}_d$ into Eq.~(\ref{Eq:AppA03}) and
collecting terms allows the equation to be rewritten as
\begin{equation}
\label{Eq:AppA06}
\left[{\vec w} - {\bf B}{\bf D}^{-1}{\vec w}_{d}\right] (1+r) +
\left[{\bf A}-{\bf B}{\bf D}^{-1}{\bf B}^{\rm T}\right]{\vec\psi} +
\left[{\vec u}-{\bf B}{\bf D}^{-1}{\vec u}_{d}\right] t
= {\vec 0}_n
.
\end{equation}
Note that since ${\bf D}$ is symmetric, so is ${\bf D}^{-1}$. 
We can also substitute ${\vec\psi}_d$ from
Eq.~(\ref{Eq:AppA03}) in for expressions obtained from
multiplying out the top and bottom rows of Eq.~(\ref{Eq:AppA01}).
This gives that the matrix equation
\begin{equation}
\label{Eq:AppA07}
\begin{pmatrix}
\xi -{\vec w}_{d}^{\rm T}{\bf D}^{-1}{\vec w}_{d}                      &
    {\vec w}^{\rm T} -{\vec w}_{d}^{\rm T}{\bf D}^{-1}{\bf B}^{\rm T}  &
    -{\vec w}_{d}^{\rm T}{\bf D}^{-1}{\vec u}_{d}                      \\
{\vec w} -{\bf B}{\bf D}^{-1}{\vec w}_{d}                              &
    {\bf A} -{\bf B}{\bf D}^{-1}{\bf B}^{\rm T}                        &
    {\vec u} -{\bf B}{\bf D}^{-1}{\vec u}_{d}                          \\
-{\vec w}_{d}^{\rm T}{\bf D}^{-1}{\vec u}_{d}                          &
    {\vec u}^{\rm T} -{\vec u}_{d}^{\rm T}{\bf D}^{-1}{\bf B}^{\rm T}  &
    \xi -{\vec u}_{d}^{\rm T}{\bf D}^{-1}{\vec u}_{d}                  \\
\end{pmatrix}
\begin{pmatrix}
1+r            \\
{\vec \psi}    \\
t            \\
\end{pmatrix}
\>=\>
\begin{pmatrix}
2 i \Im (\xi) \\
{\vec 0}_n  \\
0           \\
\end{pmatrix}
\end{equation}
has the same solutions for $r$, $t$, and ${\vec\psi}$ as does
Eq.~(\ref{Eq:AppA01}). For the cases in this paper we will not
have interactions between the input site and output site and
the $m$ sites to be decimated, so both ${\vec w}_{d}$ and
${\vec u}_{d}$ will be zero and the $(n+2)\times(n+2)$ matrix
equation to solve for $t$ becomes
\begin{equation}
\label{Eq:AppA08}
\begin{pmatrix}
\xi        \>\> & {\vec w}^{\rm T}                             & 0         \\
{\vec w}        & {\bf A} -{\bf B}{\bf D}^{-1}{\bf B}^{\rm T}  & {\vec u}  \\
0               & {\vec u}^{\rm T}                             & \xi       \\
\end{pmatrix}
\begin{pmatrix}
1+r            \\
{\vec \psi}    \\
t            \\
\end{pmatrix}
\>=\>
\begin{pmatrix}
2 i \Im (\xi) \\
{\vec 0}_n  \\
0           \\
\end{pmatrix}
.
\end{equation}
This completes the decimation RG of the $m$ sites.


\begin{figure}
   \includegraphics[width=\textwidth, bb=0bp 280bp 590bp 740bp,clip]{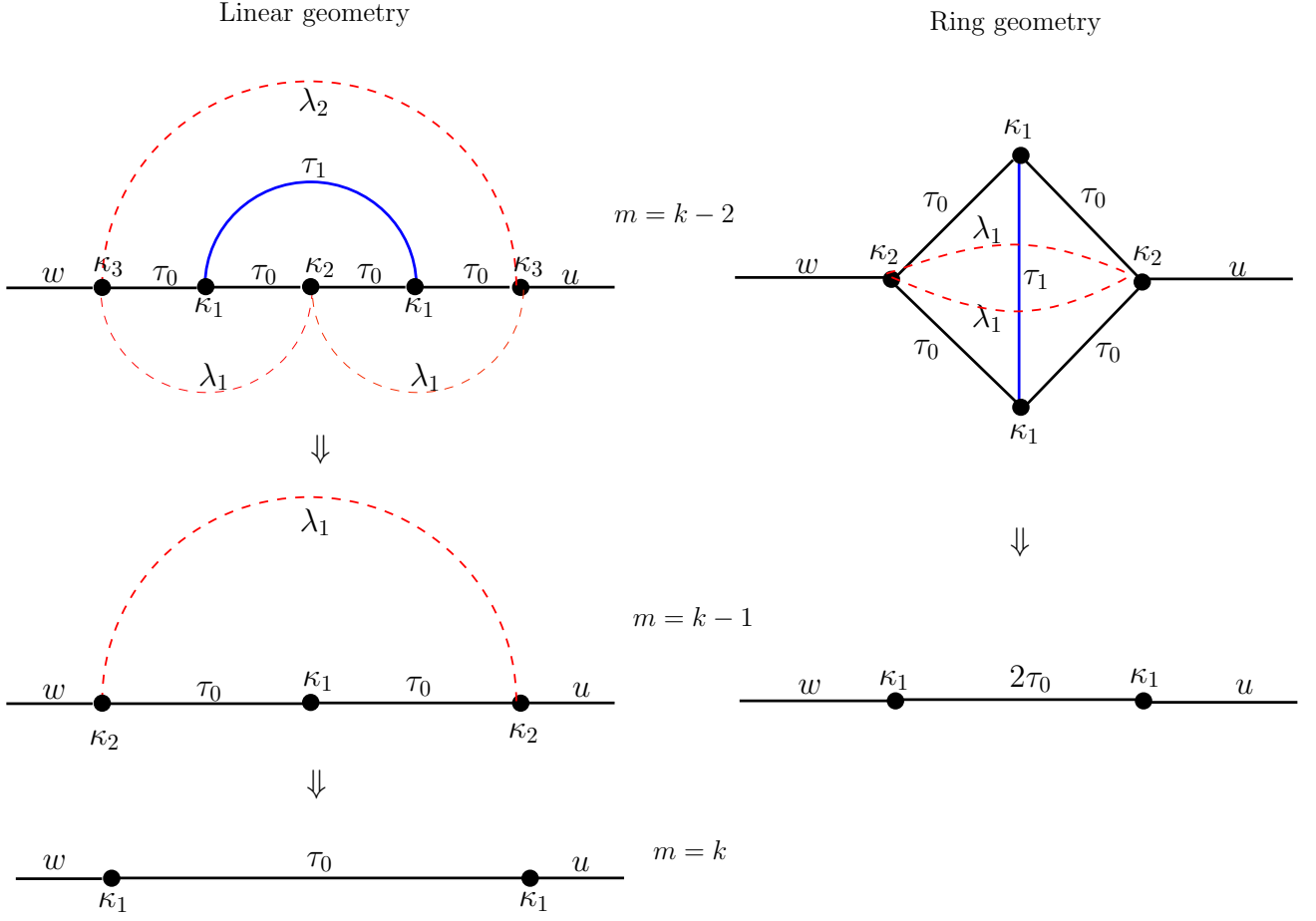}\\
  \caption{(Color online) The last steps in the RG for the linear (left) 
           and the ring (right) geometries.  In both cases one is left with 
           only two sites connected to the external leads.  Note for 
           clarity we have dropped the superscripts that denote the RG number 
           on all variables.}
  \label{EndGame}
\end{figure}

\end{widetext}

\section{Transmission from Small Renormalized Lattices \label{Sec:AppB}}

Equations(\ref{aries} -- \ref{puppy}) can be 
used to construct the $m=k-2$ state shown at the top of Fig.~\ref{EndGame}.
The subsequent RG steps require use of these equations with care. 

\subsection{Linear geometry}
For the linear geometry (left side of Fig.~\ref{EndGame}), for the $m=k-1$ RG step we 
proceed as follows. Using the RHS of Eq.~(\ref{aries}) with $i=2$ we calculate
\begin{equation}\label{}
    \tilde{\kappa}_{2}^{(k-1)} = \kappa_{1}^{(k-1)} + \kappa_{k+1} - \kappa_{k} 
     - 2(\lambda_{1}^{(k-1)} - \lambda_{k})
\end{equation}
and then substitute $\kappa_{2}^{(k-1)}$ by $\tilde{\kappa}_{2}^{(k-1)}$ in the 
RHS of Eq.~(\ref{cancer}) to get
\begin{equation}\label{}
    \kappa_{2}^{(k-1)} = \kappa_{k+1} + \lambda_{k} - \lambda_{1}^{(k-1)} 
    + \frac{1}{2}[\kappa_{1}^{(k-1)} - \kappa_{k}]
.\end{equation}

\begin{widetext}
This completes the $m=k-1$ RG step. Now for $m=k$, we proceed in a similar way. 
Using the RHS of Eq.~(\ref{kitten}) with $m=k-1$ and $\tau_k=0$, we calculate
\begin{equation}\label{}
    \tilde{\kappa}_{1}^{(k)} = \kappa_{1}^{(k-1)} + \kappa_{k+1} - \kappa_{k} 
    - 2(\lambda_{1}^{(k-1)} - \lambda_{k}) - \frac{2[\tau_{0}^{(k-1)}]^{2}}{\kappa_{1}^{(k-1)}}
\end{equation}
and then with $m=k$, $i=2$, $\lambda_{k+1} =0$ and substituting $\kappa_{1}^{(k)}$ by 
$\tilde{\kappa}_{1}^{(k)}$ in the RHS of Eq.~(\ref{aries}), we calculate
\begin{equation}\label{}
    \tilde{\kappa}_{2}^{(k)}= \tilde{\kappa}_{1}^{(k)} + \kappa_{k+2} - 
    \kappa_{k+1} - 2\lambda_{1}^{(k)}
\>.
\end{equation}
Substituting $m=k-1$ in Eq.~(\ref{puppy}) gives $\lambda_{1}^{(k)} =0$. 
Finally, with $m=k$ and substituting $\kappa_{2}^{(k)}$ by $\tilde{\kappa}_{2}^{(k)}$ in the 
RHS of Eq.~(\ref{cancer}), we get
\begin{eqnarray}\label{}
    \kappa_{1}^{(k)} &=& \kappa_{k+1} + \lambda_{k} - \lambda_{1}^{(k-1)} 
    - \frac{[\tau_{0}^{(k-1)}]^{2}}{\kappa_{1}^{(k-1)}} + \frac{1}{2}[\kappa_{1}^{(k-1)}  
    - \kappa_{k}] \nonumber \\
    &=& \kappa_{k+1} + \lambda_{k} + \tau_{0}^{(k)}- 2\lambda_{1}^{(k-1)} 
    + \frac{1}{2}[\kappa_{1}^{(k-1)}  - \kappa_{k}] \> .
\end{eqnarray}

After performing $k$ RG steps, we are left with just $1 +
2^{k-k} =2$ sites with an on-site energy corresponding to
$\kappa_{1}^{(k)}$ and an interaction of $\tau_{0}^{(k)}$
between them. One site is connected to the input by $w$ while
the other is connected to the output by $u$. In order to
decimate these two sites, we have
\begin{eqnarray}
 {\bf A} =
             \begin{pmatrix}
               \xi & 0 \\
               0 & \xi
             \end{pmatrix}
            ; \quad\quad  {\bf D} =
             \begin{pmatrix}
               \kappa_{1}^{(k)} & \tau_{0}^{(k)} \\
               \tau_{0}^{(k)} & \kappa_{1}^{(k)}
             \end{pmatrix}
           ; \quad\quad {\bf B} =
             \begin{pmatrix}
               w & 0 \\
               0 & u
             \end{pmatrix}
            \\
{\bf A'} =  {\bf A} -{\bf B}{\bf D}^{-1}{\bf B}^{\rm T} =
             \begin{pmatrix}
             \xi -\kappa_{1}^{(k)}\beta/\gamma & \beta \tau_{0}^{(k)} \\
\beta \tau_{0}^{(k)} & \xi -\kappa_{1}^{(k)}\beta\gamma
             \end{pmatrix}
\end{eqnarray}
where 
$\beta = wu/([\kappa_{1}^{(k)}]^{2} - [\tau_{0}^{(k)}]^{2})$ and $\gamma = u/w$. 
Thus after decimating all $N = 1+ 2^{k}$ sites, we have
\begin{eqnarray}
             \begin{pmatrix}
             \xi -\kappa_{1}^{(k)}\beta/\gamma & \beta \tau_{0}^{(k)} \\
            \beta \tau_{0}^{(k)} & \xi -\kappa_{1}^{(k)}\beta\gamma
             \end{pmatrix}
             \begin{pmatrix}
             1+r \\
            t
             \end{pmatrix}
  =
\begin{pmatrix}
             2i\Im(\xi)\\
0
\end{pmatrix}
.\end{eqnarray}

From the above matrix equation by taking the inverse of the $2\times 2$ matrix,
\begin{equation}
\begin{pmatrix}
1+r\\t
\end{pmatrix} 
= 
\frac{2i\Im(\xi)}{(\xi -\kappa_{1}^{(k)}\beta/\gamma)(\xi - \kappa_{1}^{(k)}\beta\gamma) - 
\beta^{2}[\tau_{0}^{(k)}]^{2}}\begin{pmatrix}
\xi -\kappa_{1}^{(k)}\beta\gamma \\ -\beta \tau_{0}^{(k)}
\end{pmatrix} \label{linear1tT}
.\end{equation}
\end{widetext}
Thus we have found the transmission $T = |t|^2$.

Now using Eq.~(\ref{Eq:AppA05}) we can calculate the wavefunctions
associated with the various sites. For simplicity we take
$w=u$. To find the wavefunction associated with the highest
level sites ($i=k+1$), we start with the $m=k$ RG step given in Fig.~\ref{EndGame}. 
Here, for Eq.~(\ref{Eq:AppA05}),  $\boldsymbol{\psi} =0$,
$
\mathbf{w_{d}} =  
\begin{pmatrix}
w \\
0 \\
\end{pmatrix}
$, 
$
\mathbf{u_{d}} =
\begin{pmatrix}
0 \\
w \\
\end{pmatrix}
$ whereas the ${\bf D}$ and ${\bf B}$ matrices are given by Eq.~(\ref{AdB}). 
Making these substitutions, we get
\begin{equation}\label{HighWave}
    \boldsymbol{\psi}_{k+1} = \frac{-w}{[\kappa_{1}^{(k)}]^{2} - [\tau_{0}^{(k)}]^{2}}
       \begin{pmatrix}
              (1+r)\kappa_{1}^{(k)} - t\tau_{0}^{(k)} \\
              t\kappa_{1}^{(k)} - (1+r)\tau_{0}^{(k)} \\
       \end{pmatrix}
.
\end{equation}
Next we can find the wavefunction associated with the immediate
lower level site ($i=k$). Here, for Eq.~(\ref{Eq:AppA05}), 
$\boldsymbol{\psi} = \boldsymbol{\psi}_{k+1}$, 
$\mathbf{w_{d}} = 0 = \mathbf{u_{d}}$, ${\bf D} =
\kappa_{1}^{(k-1)}$ and ${\bf B} = \tau_{0}^{(k-1)}
\begin{pmatrix}
    1 \\
    1 \\
\end{pmatrix}
$. Making these substitutions, we get
\begin{equation}\label{NextHighWave}
   \boldsymbol{\psi}_{k} = \frac{\tau_{0}^{(k-1)}w(1+r+t)}
   {\kappa_{1}^{(k-1)}[\kappa_{1}^{(k)}+\tau_{0}^{(k)}]}
\> .
\end{equation}
We can continue in this manner, in principle, to find the
wavefunction of all the levels below. However the calculations 
get very tedious and the expressions very complicated below 
$i=k$ and therefore are not given explicitly.

\subsection{Ring geometry}
Next consider the ring geometry (right side of Fig.~\ref{EndGame}).  
We perform $k-1$ RG steps to be left with $2^{k-(k-1)} =2$
sites. Site 1 is the nearest neighbor of site 0 and site 2.  
This makes the interaction
between the two sites to be $2\tau_{0}^{(k-1)}$.
This adds a factor of $2\lambda_{1}^{(k-1)}$ to the on-site energy of site 0. 
Thus the on-site energy of the even site 0 corresponds to 
$(\kappa_{k-1} - 2l_{k-1}) + 2l_{k-1} = \kappa_{k-1}$.
The decimation of these two sites is similar to the case of the linear geometry, 
except that we need to 
replace $\kappa_{1}^{(k)}$ by $\kappa_{1}^{(k-1)}$ and $\tau_{0}^{(k)}$ by $2\tau_{0}^{(k-1)}$. 
So $\beta = wu/([\kappa_{1}^{(k-1)}]^{2} - 4[\tau_{0}^{(k-1)}]^{2})$ and
\begin{widetext}
\begin{equation}
\begin{pmatrix}
1+r\\t
\end{pmatrix} = \frac{2i\Im(\xi)}{(\xi -\kappa_{1}^{(k-1)}\beta/\gamma)(\xi 
- \kappa_{1}^{(k-1)}\beta\gamma) - 4\beta^{2}[\tau_{0}^{(k-1)}]^{2}}\begin{pmatrix}
\xi -\kappa_{1}^{(k-1)}\beta\gamma \\ -2\beta \tau_{0}^{(k-1)}
\end{pmatrix} \label{linear2tT}
.\end{equation}
This completes the RG, giving the transmission $T =|t|^2$.

To find the wave functions here, we start at the $m=k-1$ RG
step and proceed as we did in the case of the linear geometry.
We find that 
\begin{eqnarray}
  \boldsymbol{\psi}_{k} &=& \frac{-w}{[\kappa_{1}^{(k-1)}]^{2} - 4[\tau_{0}^{(k-1)}]^{2}}
  \> 
  \begin{pmatrix}
          (1+r)\kappa_{1}^{(k-1)} - 2t\tau_{0}^{(k-1)} \\
          t\kappa_{1}^{(k-1)} - 2(1+r)\tau_{0}^{(k-1)} \\
   \end{pmatrix}
\end{eqnarray}
and
\begin{eqnarray}
\boldsymbol{\psi}_{k-1} &=& \frac{\tau_{0}^{(k-2)}w(1+r+t)}
{[\kappa_{1}^{(k-2)}+\tau_{1}^{(k-2)}][\kappa_{1}^{(k-1)}+2\tau_{0}^{(k-1)}]}
          \>
          \begin{pmatrix}
                1 \\
                1 \\
         \end{pmatrix}
.
\end{eqnarray}
\end{widetext}
Notice that the two $i =k-1$ sites are symmetric. Therefore the wavefunctions obtained above for the 
two sites are identical, as expected.  Again the wavefunction expressions for the lower 
$i$ sites become complicated upon further iteration of this methodology.  

\bibliographystyle{apsrev}

\end{document}